\documentclass[useAMS,usenatbib]{mn2e}

\usepackage{times}
\usepackage{epsfig}
\usepackage{graphicx}
\usepackage{amsmath}	
\usepackage{natbib}
\usepackage{latexsym}

\newcommand{{\kms}}{km\,s$_{-1}$}

\graphicspath{{./IMAGES}}

\def\aj{AJ}%
\def\apj{ApJ}%
\def\apjl{ApJ}%
\def\aap{A\&A}%
\def\aapr{A\&A~Rev.}%
\def\aaps{A\&AS}%
\def\mnras{MNRAS}%
\def\pasj{PASJ}%
\def\rmxaa{Rev. Mexicana Astron. Astrofis.}%
\def\nat{Nature}%
\def\gca{Geochim.~Cosmochim.~Acta}%

\begin{document}
\renewcommand{\thefootnote}{\fnsymbol{footnote}}
\title[X-ray study of the supernova remnant 3C400.2]{ A Chandra X-ray study of the mixed-morphology supernova remnant 3C400.2 
}
\author[S. Broersen \& J. Vink]{Sjors~Broersen,$^1$ Jacco~Vink$^{1,2}$  \\  
$^1$ Astronomical Institute `Anton Pannekoek', University of Amsterdam, Postbus 94249, 1090 GE Amsterdam, The Netherlands \\
$^2$ GRAPPA, University of Amsterdam, Postbus 94249, 1090 GE Amsterdam, The Netherlands }

\maketitle

\begin{abstract}

We present an analysis of archival Chandra observations of the mixed-morphology remnant 3C400.2. We analysed spectra of different parts of the remnant to observe if the plasma 
properties provide hints on the origin of the mixed-morphology class.
These remnants often show overionization, which is a sign of rapid cooling of the thermal plasma, and super-solar abundances of elements which is a sign of ejecta emission. Our analysis shows that the thermal emission of 3C400.2 can be well explained by a two component non-equilibrium ionization model, of which one component is underionized, has a high temperature ($kT\approx  3.9$~keV)
and super-solar abundances, while the other component has a much lower temperature 
($kT\approx0.14$~keV), solar abundances and shows signs of overionization. 
The temperature structure, abundance values and density contrast between the different model components suggest that the hot component comes from ejecta plasma, while the cooler component has an interstellar matter origin. This seems to be the first instance of an overionized plasma found in the outer regions of a supernova remnant, whereas the ejecta component of the inner region is underionized. 
In addition, the non-ionization equilibrium plasma component 
associated with the ejecta is confined to the central, brighter parts of the remnant, 
whereas the cooler component is present mostly in the outer regions. 
Therefore our data can most naturally be explained by an evolutionary scenario in which the outer parts of the remnant are cooling rapidly due to having swept up high density ISM, while the inner parts are very hot and cooling inefficiently due to low density of the plasma. This is also known as the relic X-ray scenario.

\end{abstract}


\section{Introduction}

An important and not well-understood class of supernova remnants (SNRs) are the so-called thermal composite or mixed-morphology remnants (MMRs, \citet{rhopetre1998, lazendicslane2006,vink2012}). These remnants are characterised by thermal X-ray emission that is centrally peaked whereas the radio emission has the familiar shell-type morphology. They have several other interesting characteristics: they are often associated with GeV gamma-ray emission \citep[e.g. ][]{giulianietal2010, uchiyamaetal2012},  there is evidence for enhanced metal abundances in some of them \citep{lazendicslane2006} although they are usually mature remnants, and they often show spectral features associated with strong cooling in the form of radiative recombination continua (RRCs) or strong He$-\alpha$/Ly$-\alpha$ X-ray line ratios of alpha-elements such as  Si,  S and Ar \citep{kawasakietal2005}.  

The centrally peaked X-ray emission of MMRs poses a problem for SNR evolution models. They typically have an age on the order of 20.000 years and are therefore expected to have a shell-like density structure with a hot, tenuous plasma in the centre, based on a Sedov evolutionary scenario. A flat interstellar matter (ISM) structure is therefore unsuited to create the centre-filled X-ray morphology that is observed, if the temperature is relatively uniform across the remnant. 
\citet*{rhopetre1998} mention two possible scenarios for the formation of the emission structure typical for MMRs, the \emph{relic X-ray emission} and the \emph{evaporating cloudlet} scenario \citep{whitelong1991}. The former scenario is the most simple scenario, in which the outer layers of the remnant have become radiative and have cooled strongly so that they hardly emit in the X-ray band, whilst the centre consists of hot plasma that has not yet cooled below $10^{6}$ K. This scenario requires a high surrounding ISM density.
\citet{whitelong1991} give a self-similar solution for a supernova that exploded in an environment where it is surrounded by a number of small, dense clouds. Due to the large filling fraction and small size of the clouds they do not alter the dynamics of the forward shock, and they survive the passage of the forward shock due to their large density. The clouds then slowly evaporate due to heating by thermal conduction, which increases the density and decreases the temperature in the centre of the remnant.
Of these two scenarios, the relic X-ray emission scenario was preferred by \citet{harrusetal1997} to explain the morphology of W44. 
One year later, \citet{coxetal1999, sheltonetal1999} used a similar scenario as \citet{harrusetal1997} to model the characteristics of the MMR W44, but added a density gradient for the ISM and thermal conduction. Thermal conduction in the model by \citet{coxetal1999} smoothes the temperature gradient from the centre to the outer layers, thereby reducing the pressure in the centre. A lower central pressure reduces the need to expand, which allows for a higher density to remain in the centre of the remnant. A problem with the scenarios using thermal conduction is that it is not a priori clear whether it can be important in SNRs, since it is strongly suppressed by magnetic fields \citep{spitzer1981, tao1995}. 

Clear evidence for overionization of thermal SNR plasmas is almost exclusively found in MMRs. Thermal plasmas in supernova remnants are often found in an underionized state, and reach ionization equilibrium on a density dependent timescale $t \simeq 10^{12.5} / n_e$ s \citep{smithhughes2010}.  Due to their ages and their frequent association with high density regions, MMRs are expected to have a plasma that is in ionization equilibrium. The fact that there is ample evidence for overionization, means the cooling rate of the plasma was faster than the recombination rate. How the rapid cooling proceeds is still unclear. Efforts have been made to localise regions in MMRs which are cooling rapidly, as an association with a high density region may mean that thermal conduction is important. For example: \citet{micelietal2010,lopezetal2013} find that the amount of cooling increases away from a molecular cloud for W49B, suggesting adiabatic cooling as the dominant cooling mechanism. \citet{uchidaetal2012} suggest the cooling timescale of the plasma cannot be explained by thermal conduction, and that adiabatic expansion is probably the dominant cause for cooling there. In addition, \citet{broersenetal2011} have shown that even at small expansion rates, simple adiabatic cooling combined with cooling through X-ray radiation can lead to a cooling rate of the plasma larger than the recombination rate. So far observational evidence for strong thermal conduction is lacking.

The ionization state of a thermal plasma can be determined by using both the electron temperature $T_{\rm e}$ and the so-called ionization temperature $T_{\rm z}$ \citep{masai1997}. The ionization temperature is the electron temperature  that one would deduce from the ionization state of X-ray emitting elements alone, which may be different from the thermodynamic temperature in case the plasma is not in ionization  equilibrium. When $T_{\rm z} <  T_{\rm e}$, the plasma is underionized, when $T_{\rm z} =  T_{\rm e}$ the plasma is in collisional ionization equilibrium (CIE) and when $T_{\rm z} > T_{\rm e}$ the plasma is overionized. There are different ways in which $T_{\rm z}$ can be determined. \citet{kawasakietal2002} determine $T_{\rm z}$ by comparing the observed Ly$-\alpha$ / He$-\alpha$ ratio of a certain element to the ratios of CIE plasmas of different temperatures. The temperature of the CIE plasma that produces the observed ratio is then $T_{\rm z}$. A different method was used by \citet{ohnishietal2011}, who determined $T_{\rm z}$ by using the CIE model in {\sc spex} to fit the plasma. This model has an additional parameter $\xi$, compared to the {\sc xspec} CIE model. This parameter can be used to simulate a non-equilibrium plasma so that $T_{\rm z} = \xi\times T_{\rm e}$. Finally there is the method used by \citet{broersenetal2011} and \citet{uchidaetal2012}, who fit a plasma using the {\sc spex} NEI model with initial temperature $kT_1 > kT_2$. 
$kT_1$ has a similar function as the ionization temperature, but it has a slightly different physical meaning as it describes the initial ionization state.  The NEI model then follows the ionization state as a function of ionization age as usual, assuming a rather sudden drop of electron temperature. 
Note that although using the He$-\alpha$ / Ly$-\alpha$ line ratios of alpha elements can show that a plasma is overionized, it has quite a large systematic error in determining the $T_{\rm z}$ \citep{lopezetal2013}. This is the result of the fact that there are emission lines of other elements which contaminate the He$-\alpha$ /Ly$-\alpha$ ratio. The strengths of the contaminating lines are $T_{\rm e}$ dependent, so that the systematic error is  $T_{\rm e}$ dependent as well. This can be taken into account by calculating the CIE line ratios including the contaminating lines.

\begin{figure}%
\includegraphics[angle=0,width=\columnwidth]{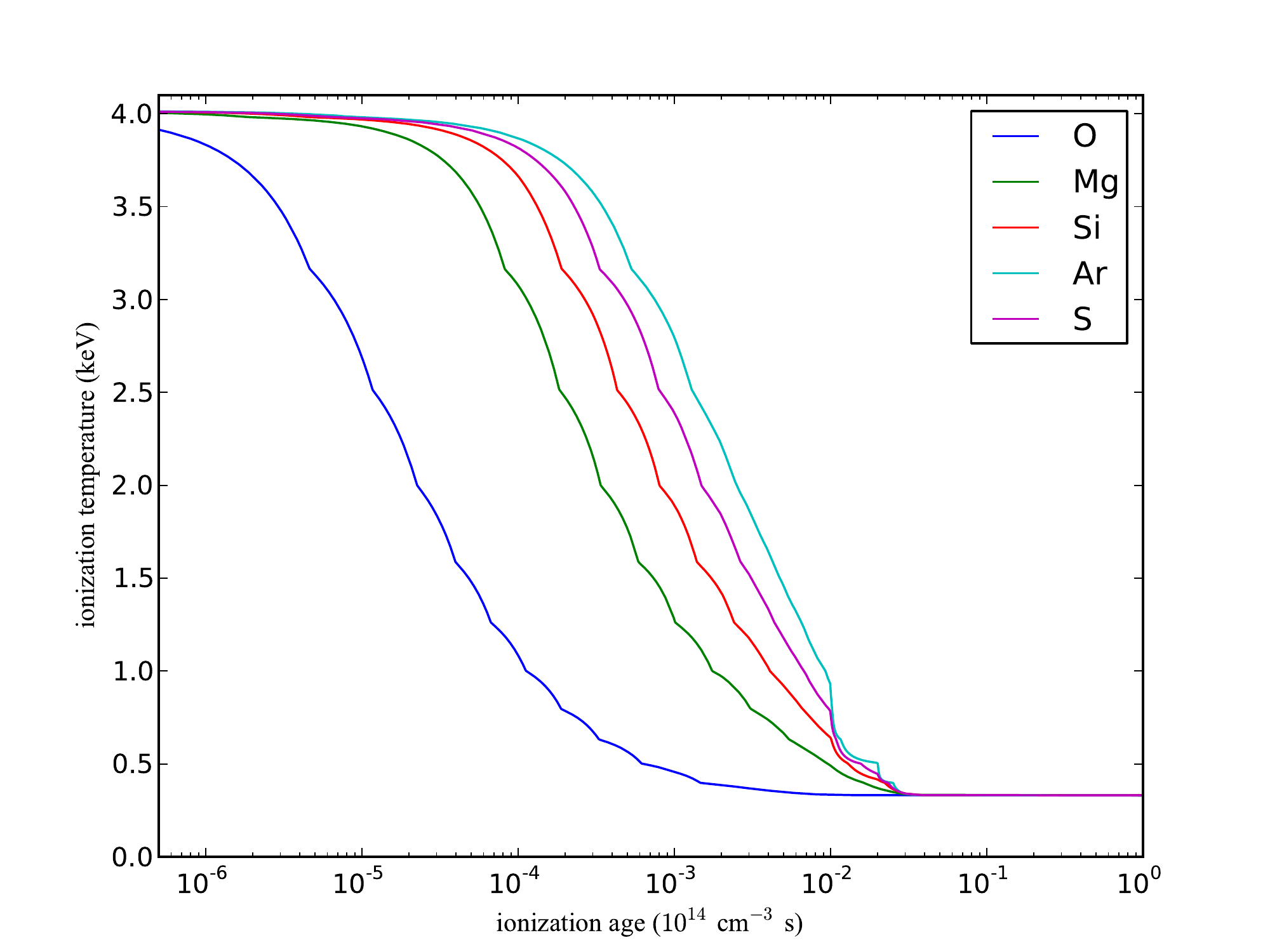} 
\caption{The ionization temperature of a plasma versus the ionization age. The NEI plasma starts at an electron and ionization temperature of 4.0 keV, 
and it evolves after the electron temperature instantly drops to 0.2 keV. 
The figure illustrates that the NEI and line-ratio/ionization-temperature  
methods for characterizing overionized plasma do not
provide identical ionization states. Moreover, the differences
depend on the element:
at certain ionization ages, different elements may have different ionization temperatures. %
  \label{fig:Tz_net}}
\end{figure}

The difference in ionization temperature between the line ratio and NEI model methods is displayed in Fig. \ref{fig:Tz_net}. This figure shows the evolution of the ionization temperature as a function of ionization age, for an NEI plasma that is cooling abruptly from 4 keV to 0.2 keV. It is clear from the figure that different elements show different ionization temperatures as the plasma evolves, so that a plasma cannot be characterised by a single ionization temperature. This also explains why different elements in SNR plasmas often show different ionization temperatures, as found by \citet[e.g.][]{lopezetal2013}. On the other hand the NEI model is not a perfect model for a cooling plasma, as it assumes the electron temperature instantly drops to a certain value. In reality the electron temperature will drop more gradual, so that the plasma evolution will be slightly different as a result of different recombination and ionization rates.

Here we present the  first analysis of a Chandra X-ray observation of 3C400.2 (also known as G53.6~--~2.2), which is an important member of the MMR class. This remnant has centrally peaked X-ray emission, as shown by Einstein IPC and ROSAT observations \citep{longetal1991, sakenetal1995}. The radio morphology can be described as two overlapping circular shells of diameters 14' and 22'  \citep{dubneretal1994}. This has led to the speculation that 3C400.2 might be two supernova remnants in contact with each other, which would make it a rare event. \citet{yoshitaetal2001} conclude however, using ASCA observations, that the remnant is the result of a single supernova explosion,  based on the similar plasma properties found in the two shells. Hydrodynamical simulations, including thermal conduction, show that the morphology of the remnant can be explained by a supernova exploding in a cavity, where the larger shell is a part of the remnant expanding into a lower density region than the smaller shell \citep{schneiteretal2006}. This is consistent with HI observations performed by \citet{giacanietal1998}, who find a denser region to the northwest of the remnant where the smaller shell is located. 
In the optical, the remnant is characterised by a shell-like structure with a smaller radius (about 8') than the radio shell \citep{winkleretal1993, ambrociocruzetal2006}. The optical emission is located in regions of low X-ray emissivity. Optical emission suggest that those parts of the remnant contain radiative shocks, and are therefore cooling efficiently. 
Distance estimates to 3C400.2 range from 2.3 -- 6.9 kpc \citep{rosado1983,milne1979,dubneretal1994,giacanietal1998}. However, the distance estimates based on radio data are obtained with the uncertain $\Sigma-D$ relationship, and the kinematic estimate of \citet{rosado1983} is based on interferograms of a small part of the remnant. We therefore consider the most recent distance estimate of $2.3\pm0.8$ kpc \citep{giacanietal1998}, which is based on HI measurements, as the more reliable one and we will use a round value of 2.5 kpc for the distance throughout this paper.

MMRs and in particular overionization of plasmas have gathered increasing attention over the past few years. In this work, we therefore aim to characterise the plasma properties of 3C400.2 in order to increase our understanding both of overionized plasmas, and the possible evolutionary scenarios of MMRs.  
We start with the data reduction and spectral analysis of different regions of the remnant, which is followed by a discussion on the measured plasma properties. We end with the conclusions. 

\begin{figure}
\includegraphics[width=\columnwidth]{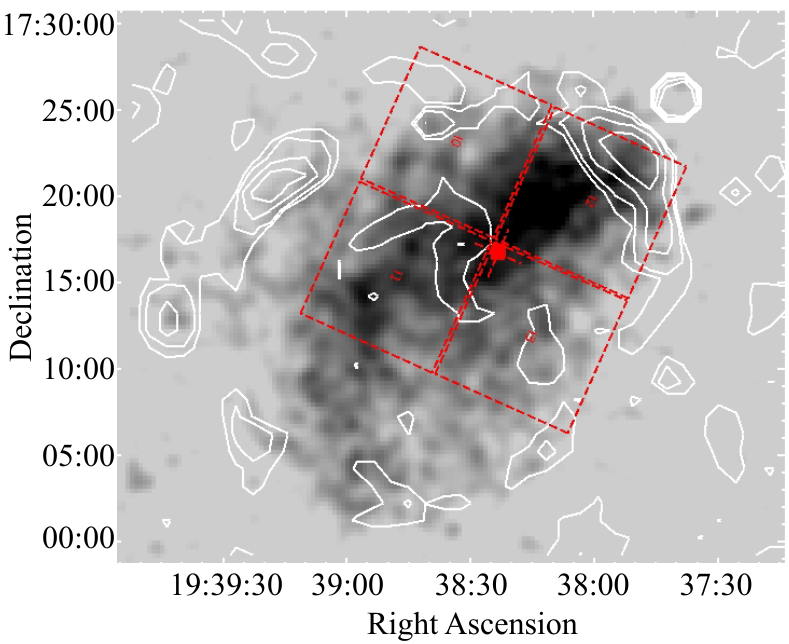}
\caption{Rosat PSPC image in an inverted grey scale, with NRAO VLA Sky Survey contours overlaid,
The Chandra ACIS-I field of view is indicated in red. }
\label{fig:rosat}
\end{figure}

\section{Data Analysis and results}

\subsection{Data reduction}
In this paper we report on the analysis of a $34$ ks archival Chandra observation (ObsID 2807) taken on August 11, 2002 with Chandra observing in imaging mode with the ACIS-I CCD array. The total number of counts in the 0.3-7.0 keV band in the full spectrum is $\sim58$k. We extracted spectra using the default tasks in the Chandra analysis software {\sc ciao} version 4.5, using the task \emph{specextract} to create spectra and weighted responses (RMF and ARF). We removed point sources from the spectral extraction regions. The spectra where grouped so that each bin contained a minimum of 15 counts. The remnant covers the whole area of the ACIS-I chips (see \ref{fig:rosat}), and there was no region available for background extraction. We therefore used the standard ACIS-I background files to create background spectra. 
We used {\sc spex} version 2.03 \citep{SPEX} for the spectral modelling. The NEI model in {\sc spex} is an extended version of the MEKAL model used in {\sc xspec} \citep{xspec}, with the added advantage that the initial temperature of the NEI model can be varied to mimic an overionized plasma.

\subsection{Spectral analysis}
\label{sec:spectral_analysis}
The Chandra ACIS-I field of view covers the part of the remnant that is mostly associated with the smaller radio shell. It includes the brightest part of the remnant in X-rays, as can be deduced from the ROSAT image in Fig.~\ref{fig:rosat}. With our spectral analysis, we aim to find differences in plasma properties between different parts of the remnant, and to investigate if this remnant conforms to the general properties of MMRs mentioned in the introduction, with regards to overionization and abundances.

We fitted the spectra using absorbed non-equilibrium ionization (NEI) models, of which the parameters are the ionization age $\tau = n_et$, the electron temperature $kT_2$, the elemental abundances and the normalisation $n_{\rm e}n_{\rm H}V$. A CIE plasma is identical to an NEI plasma when $\tau \geq 10^{12.5}$ cm$^{-3}$ s. In addition to the above listed parameters, the {\sc spex}NEI model has the initial temperature of the plasma, $kT_1$, as an optional parameter. As mentioned in the introduction putting $kT_1 > kT_2$ makes the NEI model mimic an overionized plasma, where the ionization state of the plasma is determined by $kT_1$ and $\tau$, while the continuum shape is determined by the electron temperature $kT_2 = kT_{\rm e}$. The method to check for overionization using $Ly-\alpha$ and $He-\alpha$ line strengths is not feasible for 3C400.2, since elements with isolated emission lines such as Si and S show no significant $Ly-\alpha$ lines in this remnant. 

\begin{figure}
\includegraphics[width=95mm]{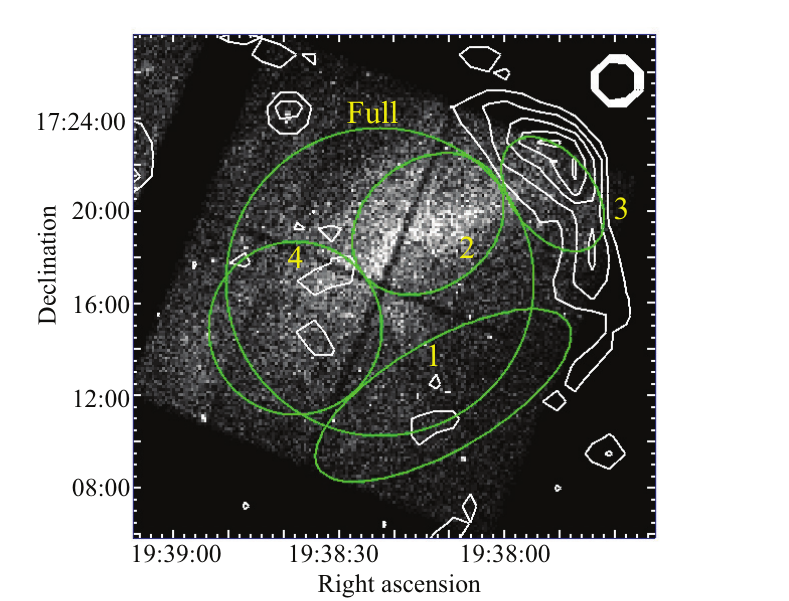} 
\caption{Chandra image in the 0.3 - 7.0 keV band with NVSS radio contours. The extraction regions of the spectra are labelled in yellow. }
\label{fig:chandra}
\end{figure}

We aimed to first find a satisfactory fit for the region covering almost the full area of the ACIS-I chips (see Fig. \ref{fig:chandra}), where we initially tried to fit the spectrum with an absorbed, single, underionized NEI model with fixed abundances, freeing abundances only when it led to a significant improvement in C-stat / degrees of freedom \citep{cash}. The $N_{\rm H}$ was always allowed to vary. A single absorbed NEI model was not sufficient to obtain an acceptable C-stat / d.o.f. in any of the extracted spectra, however. 

\begin{figure*}%
\includegraphics[angle=0]{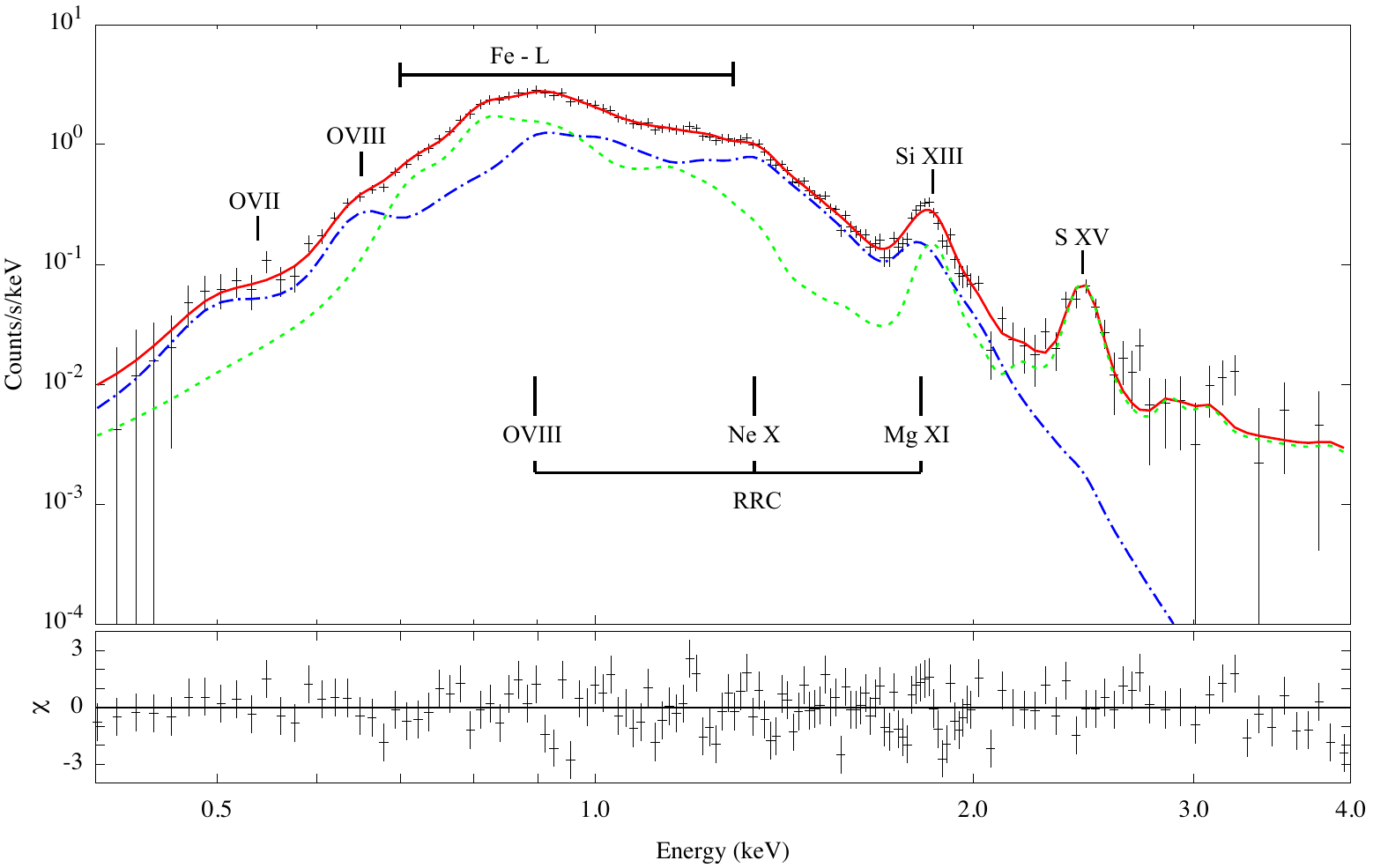} 
\caption{Spectrum of the region covering almost the whole area of the ACIS-I chips. The blue dash-dotted line shows the low $kT$ overionized NEI component, while the green dashed line shows the high $kT$ underionized NEI component. The parameters of the model are listed in Tab. \ref{tab:full_param}. Indicated in the figure are the locations of the \mbox{O\,{\sc viii}}, \mbox{Ne\,{\sc x}} and \mbox{Mg\,{\sc xi}} RRCs which are present in the overionized model (blue dash-dotted line) and the locations of important emission lines. The green dashed line represents the hot NEI component.
  \label{fig:full}}
\end{figure*}

The next step was to try to fit the spectrum with a double NEI model. Although the C-stat / d.o.f. improved significantly with respect to a fit with a single NEI model, the fit was still not acceptable. The only way the fit of the double NEI models became acceptable, was by allowing the $kT_1$ of the cooler NEI component to vary, so that $kT_1 > kT_2$, and therefore the NEI plasma is overionized. We show in detail why the overionized model fits the data better than an underionized model in section \ref{sec:overionization}. 

Fig.~\ref{fig:full} shows the spectrum of the full region. The best-fit model for this region consists of a  $kT=3.86$ keV plasma with super-solar abundances plotted as a green dashed line, combined with a $kT_2=0.14$ keV, overionized NEI component which is plotted as a blue dot-dashed line. The parameters of the model are listed in Tab. \ref{tab:full_param}. It is clear from the figure that the high $kT$ model accounts for the bulk of the Fe-L (0.7-1.2 keV), \mbox{Si\, {\sc xiii}} (1.85 keV) and \mbox{S\, {\sc xv}} (2.46 keV) line emission, while the low $kT$ model accounts for the \mbox{O\,{\sc vii-viii}} (0.56 and 0.65 keV) line emission, and continuum emission in the form of \mbox{O\,{\sc viii}}, \mbox{Ne\,{\sc x}} and \mbox{Mg\,{\sc xi}} RRCs. The super-solar abundances of Si, S and Fe in the hot NEI component suggest an ejecta origin for the plasma. 
The abundances of the high $kT$ component as well as the emission measure have quite large errors. This is caused by the fact that the continuum shape at energies larger than 2.5 keV is not well defined by the data. An ill-defined continuum strength causes large formal errors in the abundances, since the strength of the continuum and the height of the abundances are anti-correlated. Nonetheless the abundances are significantly super-solar, which means that 3C400.2 belongs to the group of MMRs with super-solar abundances. The ill-defined continuum at higher energies may be the result of the uncertainty introduced by the blank-sky background subtraction. 
Overionization is found only in the lower $kT$ plasma, and not in the high $kT$ plasma. Making the initial temperature a free parameter in the hot NEI component did not improve the fit. There is therefore no evidence for overionization in the ejecta component, which is further corroborated by the absence of strong $Ly-\alpha$ lines of Si and S. The abundances of the low $kT$ NEI component are sub-solar in the case of NE and O and solar for all other elements, which suggest a swept-up ISM origin for this plasma.   

Overall the fit is acceptable at a C-stat / d.o.f. of 257 / 227, although there are some significant residuals. The most notable residual feature is found at the position of the \mbox{Si\,{\sc xiii-xiv}} line at 1.85 keV. The model fits this line partly with the \mbox{Mg\,{\sc xi}} RRC from the cool NEI component, and partly with Si emission from the hot component. It seems that the ionization state of Si in the hot component is somewhat too high. This may be caused by the fact that this spectrum was taken from a large region of the remnant, in which in different parts the Si may be in different ionization states.
The parameters obtained from this spectral fit will be used in the discussion to determine the overall physical parameters of the remnant. 

Our best-fit model differs significantly from results obtained previously. The most recent X-ray observations of 3C400.2 were performed by \citet{yoshitaetal2001}, using the GIS instrument onboard ASCA X-ray telescope. The $N_{\rm H}$ that we find is quite similar to theirs, but they find an acceptable fit for this region using a single NEI model with $kT=0.8$ keV and $log (\tau)= 10.7-11.2$ cm$^{-3}$ s. We have tried to fit our data with a single NEI model like this, but this gives a C-stat / d.o.f.  = 606.23 / 229, even when allowing multiple abundances to vary. The differences in best-fit parameters most likely stem from the fact that the Chandra ACIS instrument has a higher spectral resolution than the GIS instrument, allowing for better constraints on plasma parameters.

\subsubsection{Overionization}
\label{sec:overionization}

\begin{figure}
\includegraphics[width=\columnwidth]{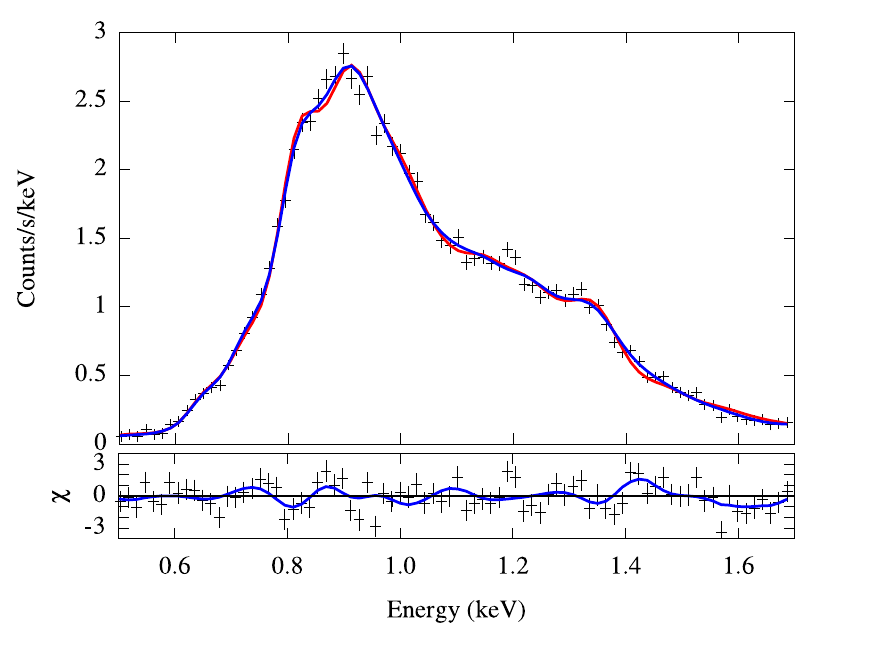}
\caption{A close up of the full spectrum, showing the effects of overionization in 3C400.2. The red solid line shows the best fit model without using cooling, while the blue line shows the best fit cooling model for this region. The bottom panel shows
the residuals of the non-cooling model with plotted in blue the difference between the cooling and non-cooling model divided by the error. This shows that the cooling model fits the spectral features 
around 0.87 keV and 1.4 keV much better than the non-cooling model.}
\label{fig:mix}
\end{figure}

Although overionization is a common feature of MMRs, it is not immediately clear that the plasma is overionized in 3C400.2, due to the lack of strong, isolated RRC features. Overionization in this remnant has a more subtle presence, which we illustrate in Fig.~\ref{fig:mix}. This figure shows a close-up of the full spectrum shown in Fig. \ref{fig:full} in the energy range 0.5 - 1.7 keV, where the red line represents the best-fit model without overionization, and the dashed blue line the best-fit overionized model. The bottom panel of the figure shows the residuals of the non-overionized model, with plotted as a blue line the two models subtracted and divided by the error. It is clear from the figure that the overionized model fits the data much better, especially in the 0.7-1.0 and the 1.3-1.7 keV region. { In the 0.5-1.7 keV range, the best-fit single absorbed NEI model has a C-stat / d.o.f. = 334/72, a double absorbed NEI model has C-stat / d.o.f. = 119/64, and the overionized model with the parameters shown in table \ref{tab:full_param} has a C-stat / d.o.f. = 92/70. The overionized model therefore fits the data significantly better than any combination of a single or double absorbed underionized NEI or CIE model. We have performed such a fitting routine where we first attempted a single underionized NEI model, then a double underionized NEI model, then allowing the initial temperature of the lower $kT$ NEI model to vary for every spectrum shown, allowing in every case the abundances to vary only if the fit improved significantly. The $N_{\rm H}$ was always allowed to vary. The C-stat / d.o.f. for the different fitting attempts are listed in table \ref{tab:spectral_parameters}. }

In the 0.7-1.0 keV energy range, the strongest emission feature is present at 0.87 keV which is not well-fit by the non-overionized model. There are two main spectral emission options which could account for the emission feature at 0.87 keV: \mbox{Fe\, {\sc xviii}} emission, which has mainly emission lines at 0.77 and 0.87 keV, and the radiative recombination continuum (RRC) of \mbox{O\, {\sc viii}}. Fe-L emission in general is often not well predicted by plasma models, as there are many different emission lines of which the strengths are not entirely known or understood \citep[e.g.][]{bernittetal2012}. Of the different ionization states of Fe which produce Fe-L emission, the Fe-L lines produced by \mbox{Fe\, {\sc xvii}} are often the most prominent in the hot plasmas found in SNRs, but they have a  weak presence in 3C400.2.  
Therefore, for a plasma to show strong \mbox{Fe\, {\sc xviii}} emission unaccompanied by \mbox{Fe\, {\sc xvii}} emission, the ionization state of the Fe plasma needs to be high, with most of the Fe in \mbox{Fe\, {\sc xviii-xx}} ionization states. We have tried to fit this feature solely with a highly ionized Fe plasma, but due to other Fe emission lines this always resulted in an unacceptable fit,
as the higher ionization states of Fe produce too much emission around 1.2 keV.
The non-overionized NEI model attempts to fit the feature at 0.87 keV using a combination of  the \mbox{Ne\, {\sc ix}} $He-\alpha$ line and \mbox{Fe\, {\sc xviii}} emission. The Ne line has a centroid of 0.92 keV, however, which results in a bad fit.
The overionized model fits this feature much better due to the addition of the \mbox{O\, {\sc viii}} RRC.
In the 1.3 - 1.7 keV range, the non-overionized model shows large residuals, where it attempts to fit continuum emission using lines of \mbox{Mg\, {\sc xi-xii}}, which has emission lines with centroids of 1.35 and 1.47 keV respectively.
The overionized model fits this region much better, using a \mbox{Ne\, {\sc x}} RRC at 1.36 keV, which improves the fit with respect to the non-overionized model. 

The above example shows that although the presence of overionization in 3C400.2 is indeed subtle, the overionized models provide a significantly better fit to the data due to the presence of RRCs. 

\subsection{Spectra}

In this section we apply a cooling model to several regions of the remnant, which show significant differences in their plasma properties. The best-fit models of the different regions are listed in Tab. \ref{tab:spectral_parameters}, where we also listed the C-stat / d.o.f. of the single NEI and double, non-cooling NEI models that we attempted to fit to the data. 

 \renewcommand\arraystretch{1.1}

\begin{table}
\caption{The best-fit model parameters of the spectrum covering the whole area of the ACIS-I chips. We used the solar abundances from \citet{abundances}. Note that the hot ejecta plasma is \emph{under}ionized, while the ISM component is \emph{over}ionized. 
The errors represent the 1$\sigma$ confidence interval. }

\begin{tabular}{llll}
Component &  Parameter & Unit & value \\ 
\hline
Ejecta & $N_{\rm H}$ &$10^{21}$ cm$^{-2}$& $6.08^{+0.15}_{-0.12}$ \\ 
& $n_{\rm e}n_{\rm H}V$&$10^{55}$ cm$^{-3}$& $2.31^{+0.83}_{-0.67}$ \\ 
&$kT_{\rm e}$ && $3.86^{+0.30}_{-0.28}$ \\ 
&$\tau$ &$10^{10}$ cm$^{-3}$ & $2.02^{+0.05}_{-0.06}$ \\ 
&Si && $3.11^{+1.23}_{-0.84}$ \\ 
&S && $6.09^{+2.73}_{-1.80}$ \\ 
&Fe && $16.6^{+7.2}_{-4.8}$ \\ 
& Luminosity &$10^{31}$erg s$^{-1}$ & $68$\\
\hline
ISM &$n_{\rm e}n_{\rm H}V$ &$10^{58}$ cm$^{-3}$& $1.17^{+0.21}_{-15}$ \\ 
&$kT_{\rm 1}$ && $0.42^{+0.03}_{-0.02}$ \\ 
&$kT_{\rm e}$ && $0.14^{+0.01}_{-0.01}$ \\ 
&$\tau$ &$10^{10}$ cm$^{-3}$ s& $26.4^{+7.8}_{-6.03}$ \\ 
&Ne && $0.40^{+0.04}_{-0.04}$ \\ 
& Luminosity & $10^{31}$erg s$^{-1}$ & $4.5$ \\
\hline
&C-stat / d.o.f.&				&  257.84 / 228 \\

\end{tabular}
\label{tab:full_param}
\end{table}

\subsubsection{Region 1: southern part of the remnant}

\begin{figure}
\includegraphics[width=84mm]{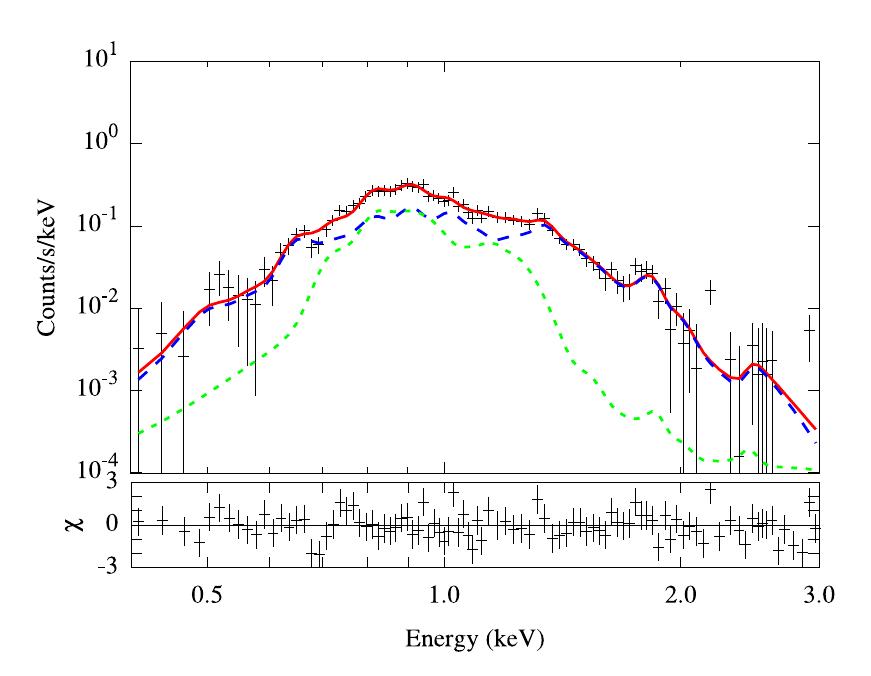}
\caption{The spectrum of region 1, fitted with the model of which the parameters are listed in Tab. \ref{tab:region1_alt_param}. The red solid line shows the complete model. The blue long dashed line shows the low $kT$ NEI component while the green short-dashed line shows the high $kT$ component. }
\label{fig:reg1_alt_spectrum}
\end{figure}

This spectral extraction region coincides with the part of the remnant that is weak in X-ray emission, as is apparent from Fig.~\ref{fig:chandra}, but is the brightest in terms of optical emission \citep{winkleretal1993}. This means that these parts of the remnant either show radiative shocks, where the shock velocity is lower than ~200 km~s$^{-1}$ for which the post-shock plasma cools efficiently, or the plasma has otherwise cooled to below $10^{6}$ K. 
This region has two models which fit nearly equally well. The first model (C-stat / d.o.f. = 81 / 87), of which the parameters are shown in Tab. \ref{tab:region1_alt_param}, has similar properties to the model for the full region. It has an ISM component with $kT_{\rm e} < kT_1$ and an ejecta component with $kT_{\rm e} = 3.1$ keV. Compared to the full model, however, the ejecta component contains no overabundance of Si. The feature in the spectrum around 1.8 keV is fit by the Mg IX RRC of the cooling component. The $\tau = 3.20^{+0.89}_{-0.68}\times10^{10}$ cm$^{-3}$ s combined with the high temperature produces a Si line with a centroid at $\sim 1.82$ keV, which is too high to fit the feature. 
In addition, the Fe abundance is very poorly constrained: when left as a free parameter the best fit value raises to $\sim250$ times solar abundance, suggesting a plasma consisting purely of Fe. This is not realistic for 3C400.2, especially when considering the best fit values of the other regions, and therefore we fixed the Fe abundance at 15, close to the value obtained for the full region. The spectrum with the best-fit model is shown in Fig. \ref{fig:reg1_alt_spectrum}.  

\begin{table}
\caption{Parameters of the alternative model for region 1. The errors represent the 1$\sigma$ confidence interval. }
\label{tab:region1_alt_param}
\begin{tabular}{llll}
Component &  Parameter & Unit & value \\ 
\hline
Ejecta & $N_{\rm H}$ & $10^{21}$ cm$^{-2}$&  $5.61^{+0.47}_{-0.45}$ \\
&$n_{\rm e}n_{\rm H}V$ &$10^{54}$ cm$^{-3}$& $2.59^{+0.28}_{-0.26}$ \\ 
&$kT_{\rm e}$ && $3.20^{+0.89}_{-0.68}$ \\ 
&$\tau$ &$10^{10}$ cm$^{-3}$ s& $2.12^{+0.08}_{-0.08}$ \\ 
&Fe && $015$ (fixed) \\ 
& Luminosity & $10^{31}$erg s$^{-1}$ & $4.2$ \\
\hline
 ISM & $n_{\rm e}n_{\rm H}V$&$10^{56}$ cm$^{-3}$& $7.49^{+2.13}_{-1.41}$ \\ 
&$kT_{\rm 1}$ & keV & $0.77^{+0.27}_{-0.15}$ \\ 
&$kT_{\rm e}$ & keV & $0.19^{+0.01}_{0.01}$\\
&$\tau$ &$10^{10}$ cm$^{-3}$ & $86.5^{+24.9}_{-21.8}$ \\ 
& Luminosity &$10^{31}$erg s$^{-1}$ & $1.6$\\

\hline
&C-stat / d.o.f.&				&  81 / 87 \\

\end{tabular}

\end{table}

The second model for this region (C-stat / d.o.f. = 78.2 / 84) consists of two cooling NEI components of which the parameters are listed in Tab. \ref{tab:spectral_parameters}. The coolest component has $kT_{\rm e} = 0.06$ keV, the lowest electron temperature of all regions in the remnant. The spectral features around 0.8-0.9 keV (see Fig. \ref{fig:spectra}) are in this model fit by a combination of \mbox{O\, {\sc vii-viii}} RRCs at 0.74 and 0.87 keV. In addition it shows a \mbox{Ne\, {\sc ix} } RRC at 1.2 keV. There is virtually no line emission present in this component. The hotter component with $kT_{\rm e}=0.53$ keV shows the \mbox{O\, \sc{viii}} emission line coupled with an RRC, but otherwise again very little line emission. The abundances of both components are mostly solar, with only O being overabundant in the hotter component. 
Contrary to the full spectrum and spectra of other extraction regions, there is no super solar abundance of Si, S or Fe present in the hot component for this region. The Fe abundance is sub-solar in the hotter component, while in the cooler component the temperatures are such that Fe has an ionization state $<$\mbox{Fe\,{\sc xvii}}, which does not show significant emission in the X-ray band. 
The fact that there is practically no line emission from elements with higher mass than Mg is not unexpected in a plasma that is cooling rapidly, since the recombination and ionization rates of an element depend strongly on its charge. 

The two models are statically almost equally probable. The first model fits the 0.8-0.9 keV features slightly better, while the second model fits the data better in the 0.4-0.8 keV region. Based on the abundances, especially the absence of Si and the high over abundance of Fe, we deem the double cooling NEI model slightly more realistic. The strong cooling fits well with the spatial coincidence with the optical emitting region in the remnant. 

\subsubsection{Region 2: brightest part of the remnant}

\begin{figure*}%
\begin{center}$
\begin{array}{cc}
\includegraphics[trim=0 0 0 0,clip=true,width=0.5\textwidth,angle=0]{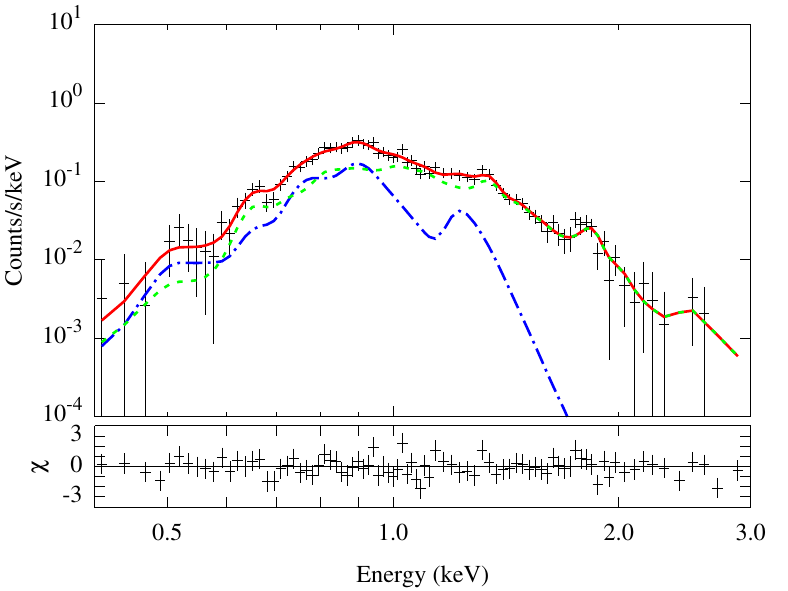} &
\includegraphics[trim=0 0 0 0,clip=true,width=0.5\textwidth,angle=0]{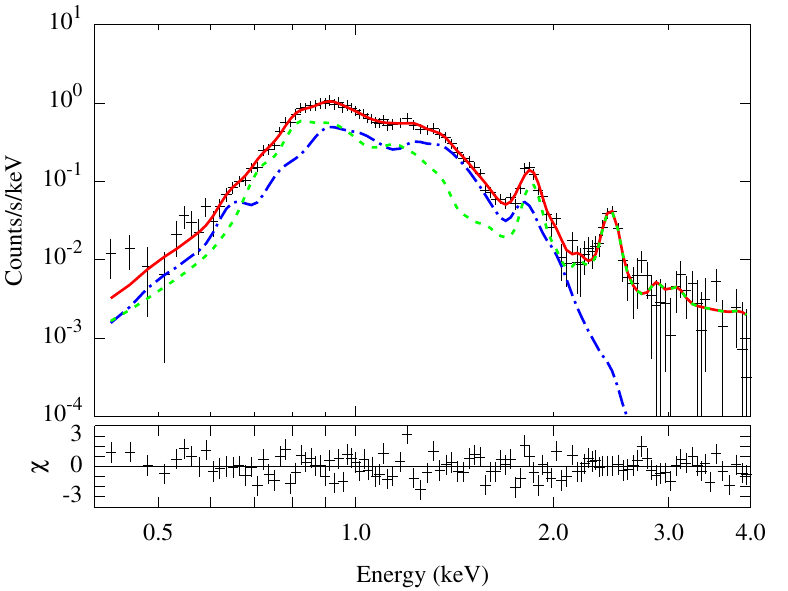} \\
\includegraphics[trim=0 0 0 0,clip=true,width=0.5\textwidth,angle=0]{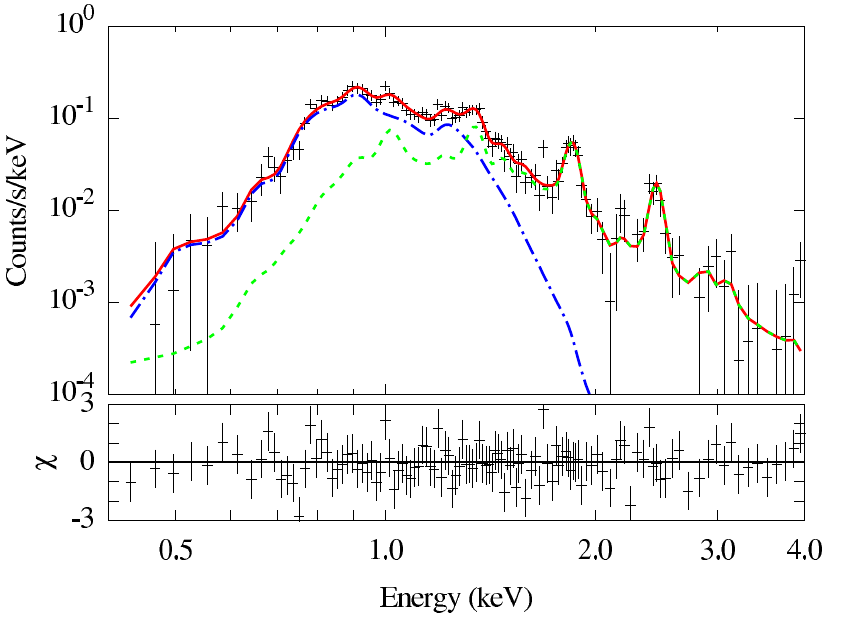} &
\includegraphics[trim=0 0 0 0,clip=true,width=0.5\textwidth,angle=0]{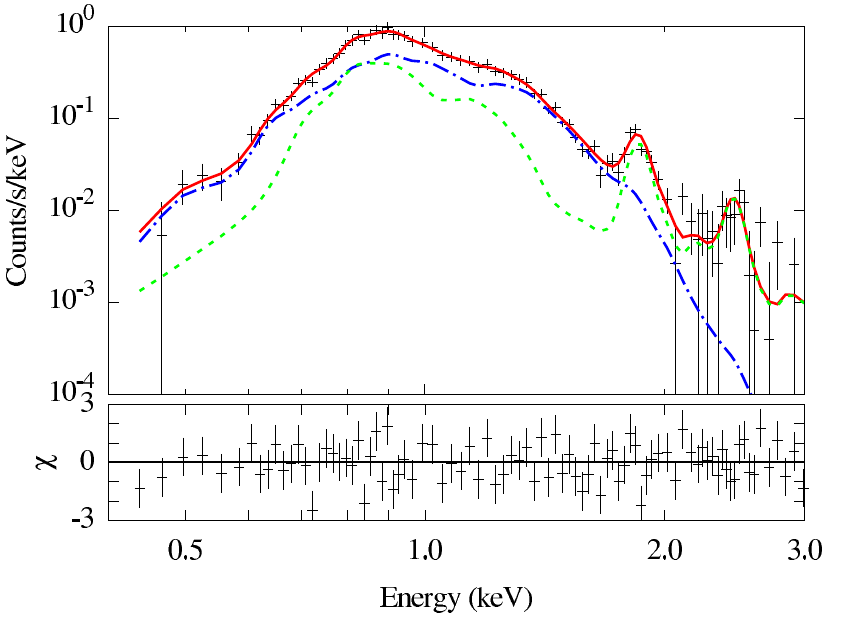}  \\
\end{array}$
\end{center}
\caption{From left to right, top to bottom: spectra of region 1, region 2, region 3 and region 4. The models are described in the text. In general the green dashed line represents the cooler NEI component, while the dash-dotted blue line represent the hotter NEI component. The total model is plotted as a solid red line. Note that  
  \label{fig:spectra}}
\end{figure*}

This region was extracted from the brightest part of the remnant in X-rays (see Fig.~\ref{fig:chandra}). The spectrum is shown in Fig.~\ref{fig:spectra}, top right. The best fit model again contains a hot ($kT_{\rm e}=3.59$ keV) underionized ($\tau=1.87\times10^4$ cm$^{-3}$ s) NEI component with super-solar abundances, plotted as a green dashed line, and a cooler ($kT_{\rm e}=0.63$ keV) NEI component that is overionized, plotted as a blue dot-dashed line. As expected, the best fit model is very similar to the model of the full region, since the full spectrum is dominated by emission from the brightest part of the remnant. The plasma parameters of the hot component are identical to the full region within the errors, but the $N_{\rm H}$ is significantly higher at $7.62^{+0.26}_{-0.24}\times10^{21}$ cm$^{-2}$, compared to $6.08^{+0.15}_{-0.12}\times10^{21}$ cm$^{-2}$. In addition, the initial temperature $kT_{\rm 1}=0.63$ keV for the cooler model is somewhat higher in this region than for the central region ($kT_1$=0.42), while the electron temperatures are identical within the errors. In general the electron temperatures for the cool component are very similar throughout the remnant, while the initial temperature shows significant variation. 

\renewcommand\arraystretch{1.1}

\begin{table*}
\centering
\begin{minipage}{126mm}
\caption{Best-fit model parameters for the spectral regions shown in Fig.~\ref{fig:rosat}. The abundances of the unlisted elements are fixed at their solar values.  
The errors represent a $1\sigma$ confidence interval. The C-stat /  d.o.f. of the single NEI and double, non-cooling NEI models we attempted to fit to the data are listed.}
\label{tab:spectral_parameters}
 \renewcommand{\tabcolsep}{.1cm}
\renewcommand{\arraystretch}{1.2}
\begin{tabular}{llllll}
\rule{0pt}{5mm}
&  & \multicolumn{4}{c}{Region} \\
\cline{3-6}

 Parameter & Unit &  1 &    2 &   3 &   4\\ 
\vspace{-5mm} \\
\hline
$N_{\rm H}$&$10^{21}$ cm$^{-2}$		& $4.76_{-0.88}^{+0.45}$ 	&    $7.62_{-0.24}^{+0.26}$ &	     $9.72_{-1.32}^{+0.53}$ &	      $5.49_{-0.23}^{+0.23}$      \\
$n_en_{\rm H}V$&	$10^{58}$ cm$^{-3}$& $0.03_{-0.01}^{+0.01}$ 	&    $1.79_{-0.40}^{+0.42}(\times10^{-3})$ &	     $8.80_{-2.70}^{+0.59}(\times10^{-3})$ &	      $0.275_{-0.24}^{+0.42}(\times10^{-3})$ 	\\
$kT_{\rm 1}$	& keV 				&$0.53_{-0.04}^{+0.04}$ 	&$-$ &$-$ & $-$		\\
$kT_{\rm e}$	& keV				& $0.25_{-0.02}^{+0.04}$ 	&    $3.59_{-0.39}^{+0.44}$ &	     $0.71_{-0.13}^{+0.06}$ &	      $3.23_{-0.72}^{+0.67}$ 	\\
  $\tau$ &	$10^{10}$ cm$^{-3}$ s		& $ 1.87_{-1.87}^{+2.77}$   	&    $2.03_{-0.06}^{+0.06}$ &	       $19.58_{-1.08}^{+2.55}$ &		$1.94_{-0.14}^{+0.17}$	  \\
  O 	&							& $2.67_{-0.61}^{+0.91}$		&$-$ &$-$ &$-$ \\
  Ne &							&$<0.24$		&$-$  &$-$ & $-$ \\		
  Si&								& 	$-$  					&    $2.75_{-0.57}^{+0.81}$ &  	   $1.34_{-0.18}^{+0.27}$ &  	    $9.53_{-5.75}^{+25.05}$        \\
  S&								& 	$-$ 					&    $5.05_{-1.09}^{+1.62}$ &  	   $2.98_{-0.56}^{+1.22}$ &  	    $12.2_{-7.9}^{+60.8}$        \\
  Fe&								&	 $0.62_{-0.24}^{+0.20}$  	&    $12.8_{-2.8}^{+6.6}$ & 	   $0.47_{-0.47}^{+0.28}$ &  	    $31.8_{-4.5}^{+78.2}$       \\
  Luminosity & 		erg s$^{-1}$		& $2.7\times10^{31}$ &$4.1\times10^{32}$ & $9.53\times10^{31}$ & $1.4\times10^{32}$\\  
  \hline
$n_en_{\rm H}V$&	$10^{58}$ cm$^{-3}$& $0.16_{-0.06}^{+0.15}$    	&    $0.31_{-0.06}^{+0.07}$ &  	   $2.08_{-1.24}^{+0.13}$ &  	    $0.29_{-0.05}^{+0.07}$	     \\
  $kT_{\rm 1}$ & keV 				&  $0.21_{-0.02}^{+0.08}$ 	& $0.63_{-0.07}^{+0.11}$ & $0.15_{-0.01}^{+0.01}$ & $0.32_{-0.04}^{+0.03}$ \\
  $kT_{\rm e}$&keV					& 	$0.06_{-0.01}^{+0.01}$    	&    $0.11_{-0.01}^{+0.01}$ &  	   $0.10_{-0.01}^{+0.01}$ &  	    $0.14_{-0.01}^{+0.02}$	     	 \\
  $\tau$&	cm$^{-3}$ s				& 	$9.40_{-9.40}^{+70.0}(\times10^{10})$    	&    $5.49_{-0.87}^{+0.89}(\times10^{11})$ &  	   $2.18_{-2.18}^{+0.52}(\times10^{8})$ & $4.63_{-4.63}^{+7.62}(\times10^{10})     $ 	\\
  O & 							& $-$ 	  &    $3.08_{-0.49}^{+0.53}$&$-$ &$-$ \\	
  Ne&							& 	 $-$     &    $-$ & $2.07_{-0.51}^{+0.25}$ & $0.59_{-0.15}^{+0.35}$  \\  				       
Luminosity &  erg s$^{-1}$			& $5.3\times10^{26}$&$1.6\times10^{31}$ &$ 7.2\times10^{28}$ & $7.5\times10^{30}$ \\ 
\hline
C-stat / d.o.f. &				&  	78.2 / 	84		& 118.40 /103			& 93.68  / 108	& 90.57 / 90\\
\hline
C-stat / d.o.f. single NEI & &140.5 / 87 & 226 / 106 & 135 / 110 & 253 / 93 \\
C-stat / d.o.f. double NEI & &115 / 86 & 163 / 102 & 120 / 107 & 149 / 88 \\
\end{tabular}
\end{minipage}
\end{table*}

\subsubsection{Region 3: The northwestern part of the remnant}

This region was taken from the utmost NW part of the Chandra FOV, which overlaps with the radio shell. The spectrum is shown in Fig.~\ref{fig:spectra} bottom left. The hydrogen column of the best-fit model of this region is  $9.7\times10^{21}$ cm$^{-2}$. Although equal within the errors to the $N_{\rm H}$ of region 2, it is significantly higher than the rest of the remnant. An increasing $N_{\rm H}$ towards the NW of the remnant is consistent with the notion of the remnant expanding into a dense ISM cloud in the NW, where it is expected that the emission from the outer parts of the remnant travel through a larger amount of material than the emission from the inner parts. Significant variation in foreground absorption cannot be ruled out, however.

The best-fit model again has a somewhat hotter, underionized NEI component with super-solar abundances combined with a cooler, overionized NEI component. However, the electron temperature of the hotter component is significantly smaller at $kT_e = 0.71$ keV, than those of the full spectrum and region 2 ($\approx 3.75$ keV), while the $\tau$ is significantly larger at $\sim2\times10^{11}$ cm$^{-3}$ s, compared to $\tau=2\times10^4$ cm$^{-3}$ s. And while the abundances are significantly super solar for Si and S, the Fe abundance is lower. It seems, therefore, that the Fe emission is confined to the central parts of the remnant. 

\subsubsection{Region 4: The centre of the remnant}

Region 4 was taken from the SW part of the Chandra FOV, which is situated roughly in the centre of the total remnant, as can be deduced from the ROSAT image. The spectrum of this region has again similar parameters as the spectra from the full remnant and region 2. It is plotted in Fig.~\ref{fig:spectra} bottom right, where the hot NEI component is plotted in a green dashed line, while the blue dot-dashed line represents the cooler NEI component. The $N_{\rm H}=5.49\times10^{21}$ cm$^{-2}$ of the best-fit model for this region confirms the trend of increasing $N_{\rm H}$ towards the NW part of the remnant. The plasma parameters of the hot component are equal within the errors to the temperatures of the full spectrum and the spectrum of region 2. Overall the model for this region confirms the notion of hot ejecta being confined to the central, brighter part of the remnant. 

\section{Discussion and Conclusion}

From our spectral modelling we obtain the following. The overall spectrum is well-fitted by a two component NEI model plasma, of which one is \emph{under}ionized with a high $0.71<kT<3.86$ keV and super-solar abundances of Si, S and Fe, while the other NEI component has a lower $0.06<kT_{\rm e}<0.15$ keV, is \emph{over}ionized and has approximately solar abundances. The central parts of the remnant show significantly higher abundances than the outer parts, which is apparent in region 1, where no super-solar abundances are found.
Region 1 coincides with the optically emitting region, and shows the lowest electron temperature. 
Although different parts of the remnant show slightly different plasma properties in terms of initial temperature $kT_{\rm 1}$ and electron temperature $T_{\rm e}$, the parameters are consistent with a hot ejecta plasma confined to the central part of the remnant, which is surrounded by a rapidly cooling, overionized swept up ISM plasma. 

\subsection{Shocked mass}

Using the parameters listed in Tab. \ref{tab:full_param}, we can estimate of the amount of X-ray emitting mass, both in ejecta and ISM, present in the SNR. The emission measure of the cool component is $1.2\pm0.2\times10^{58}$ cm$^{-3}$. We use a distance of 2.5 kpc and a spherical volume of the emitting region. The spectral extraction region has a radius of 6.67 arcmin (4.85 pc), which corresponds to $V_{\rm total} = 1.4\times10^{58}d_{2.5}^{3}$ cm$^{3}$. If we assume that the hot and the cool component are two separate plasmas, which both occupy part of the total emitting volume and which are in pressure equilibrium, we can get a unique solution for the density and shocked mass of the different components. 
The reason is that if $P_{\rm hot}$ = $P_{\rm cool}$ then also $n_{hot}kT_{hot} = n_{cool}kT_{cool}$, where $n$ is the number density. For $n_{\rm e}$ = 1.2 n$_{\rm H}$, the number density $n_{\rm H}$ of a component is given by $(EM / 1.2 / V )^{1/2}$, so that:

\begin{equation}
\left(\frac{EM_{\rm cool}}{1.2 V_{\rm total}(x)}\right)^{1/2} kT_{\rm cool} = \left(\frac{EM_{\rm hot}}{1.2 V_{\rm total}(1-x)}\right)^{1/2} kT_{\rm hot},
\end{equation} 
where EM = $n_{\rm e}n_{\rm H}V$, $V_{\rm hot} = (1-x)V_{\rm total}$ and $V_{\rm cool}=xV_{\rm total}$. 
The above equation is equal for x = 0.4, so that $n_{\rm cool}=~1.3(d_{2.5})^{-1/2}$ cm$^{-3}$ and $n_{\rm hot}~=~0.05(d_{2.5})^{-1/2}$~cm$^{-3}$. The respective masses are $M_{\rm cool} = 8.7(d_{2.5})^{5/2}$~M$_{\odot}$ and $M_{\rm hot}~=~0.46(d_{2.5})^{5/2}$~M$_{\odot}$.

The total emitting volume is uncertain, since the line of sight depth of the remnant is unknown and therefore the emitting volume might be a factor of two greater. 
Note that the total mass in the remnant is larger than we calculate here, since the Chandra FOV covers about half of the total area of 3C400.2, and we only estimate the mass of the plasma
with $T>10^6$~K.

The total mass of ejecta and ISM is quite similar to the mass found by \citet{yoshitaetal2001}, who found a mass of $6.7\pm 1.2$~M$_{\odot}d_{2.5}^{5/2}$ 
for the whole remnant. 
This is surprisingly small for a mature remnant like 3C400.2, even
if the remnant would be located at twice the assumed distance.
However, it should be noted that this is only the X-ray emitting mass, and 
most of the mass may in fact be `hiding'
in the plasma cooled below $10^6$~K.
The hot component, with super-solar abundances, only makes up a small fraction
of the mass, and the total ejecta mass seems very low if it is the ejecta component of a massive star.
However, here it should be noted that
only the inner regions of a massive star have enhanced metallicity. 
The total metallicity is a function of stellar mass, with stars around
13~M$_\odot$ producing only 0.3~M$_\odot$ of oxygen \citep{vink2012}.
Therefore a low mass of ejecta-rich material is consistent with a relatively low
mass for the exploding massive star. 
This does, however, also suggest that
the cooler plasma, which we designated ISM, may partially or completely
consist of the hydrogen-rich envelope of the star.
An explanation for why, in particular, ejecta material remains hot needs to be addressed
by detailed hydrodynamical simulations, which should incorporate the
effects of the stellar wind of the progenitor.
                                   
The low mass of the metal-rich component is also consistent with a Type Ia
origin for the remnant. In that case the cooler component could be solely shocked
ISM. The enhanced iron abundances may indeed hint at a Type Ia origin, although the total mass of Fe at an abundance of 15 times solar is still much lower than the H mass.
In addition, the association with the HI regions in the NW part of the remnant makes a core-collapse origin more likely.
To settle on the origin of the remnant, it would be important to reconfirm
the distance estimate of 2.5~kpc by \citet{giacanietal1998}, as a larger distance
estimate would favour a core collapse origin. It would also be helpful to identify
a stellar remnant in 3C400.2. The Chandra image does not show any evidence
for a bright point source that could be the cooling neutron star.

\subsection{Evolutionary scenario}

The above densities suggest a hot, metal enriched, tenuous plasma surrounded by a dense, cooler plasma which is cooling rapidly. This is consistent with the shell-like density structure expected from a Sedov evolutionary scenario. This is not the only MMR remnant in which such a temperature and density gradient is observed as \citet{kawasakietal2002} also find a two temperature best-fit model and overionization  in IC 443. They find that the central ejecta region has a temperature of 1 keV, compared to 0.2 keV in the outer layers. However, overionization is present only in the central ejecta rather than in the outer layers, while we find overionization in the outer layers and not in the ejecta. 

In the introduction we mentioned three different evolutionary scenarios that might explain the centrally peaked X-ray emission: the evaporating cloudlet scenario \citep{whitelong1991}, the \citet{coxetal1999} scenario with high surrounding density including thermal conduction, and the relic X-ray emission scenario. Our observations show that the plasma is best explained by a low density, hot interior surrounded by a high density, lower temperature plasma. The total X-ray emitting mass is relatively low, for a mature remnant. But this
is likely an indication that most of the remnant mass is not emitting in X-rays.
In addition, overionization is only present in the cool plasma, suggesting that it is cooling more efficiently than the hot plasma. Our results are perhaps most naturally explained by the simplest evolutionary scenario of the three: the relic X-ray emission scenario \citep{rhopetre1998}.  As a result of the high surrounding ISM density, the outer layers have cooled below temperatures capable of emitting in X-rays and the interior is still hot but has a low density and is therefore not cooling efficiently. We cannot rule out the presence of thermal conduction, but we do not find evidence for overionization as a result of rapid cooling for the hot centrally confined plasma. 
Moreover, the X-ray emitting plasma inside the remnant is clearly not isothermal,
as indicated by the model of  \citet{coxetal1999}.
As a final note, thermal conduction has mainly been introduced to models explaining the 
evolutionary scenarios of MMRs based on the then current observations of generally lower 
spectral and spatial resolution, which showed little to no temperature gradient in the 
remnants. 
However, more modeling is needed to understand whether local density alone determines
if a remnant will evolve into a MMR, or whether some other conditions,
such as pre-supernova evolution or ejecta structure, are important as well.

The overionization of thermal plasmas can quite naturally occur in MMRs. The high initial ISM density allows the plasma to reach CIE on a timescale smaller than the age of the remnant, after which a combination of adiabatic and radiative cooling can make the cooling rate of the plasma higher than the recombination rate, as shown in \citet{broersenetal2011}. The higher surrounding ISM density of MMRs might then be the determining factor for the occurrence of overionization compared to non-MMRs, as already noted in \citet{vinkreview}. Indeed, all remnants cool adiabatically and by radiation, but not all remnants expand in a high enough ISM density to reach CIE and then overshoot to overionization.   

\section{Summary}

We have analysed an archival Chandra observation of the mixed morphology remnant 3C400.2. Our results can be summarised as follows:
\begin{itemize}
\item The plasma of the mixed-morphology SNR 3C400.2 is best fitted by a combination of a hot, underionized plasma with low density, and a cooler, overionized plasma with high density. To our knowledge, this is the first evidence for a combination of an overionized outer shell surrounding an ejecta-rich, underionized inner region in an MMR. 
\item The hot plasma shows significant overabundances of Fe, Si and S, suggesting an ejecta origin, with Fe enhanced in the central part.
\item Overionization is significantly present in all parts of the remnant covered by the Chandra field of view.
\item The X-ray emitting masses of the plasma components are $M_{\rm cool} = 8.7(d_{2.5})^{5/2}$~M$_{\odot}$ and $M_{\rm hot} = 0.46(d_{2.5})^{5/2}$~M$_{\odot}$. 
\item This low overall mass suggests that most of the X-ray emitting mass is from mix of metal-rich and
hydrogen-rich (envelope) ejecta from a not too massive core collapse supernova, or the
remnant has a Type Ia origin.
\item The observations are best explained by a scenario in which the centrally peaked X-ray emission is caused by a hot, metal enriched, tenuous plasma. Due to the high surrounding ISM density the outer parts of the remnant have cooled efficiently towards a temperature below which they do not radiate in observable X-ray emission.
\item The overionization can be naturally explained by efficient cooling due to a high ISM density in combination with adiabatic expansion.
\end{itemize}

\section{Acknowledgements}

The authors would like to thank the anonymous referee for helpful comments and suggestions. The scientific results reported in this article are based on data obtained from the Chandra Data Archive. We also made use of the ROSAT and NVSS archives.


\begin{thebibliography}{}

\bibitem[\protect\citeauthoryear{{Ambrocio-Cruz}, {Rosado} \& {de La
  Fuente}}{{Ambrocio-Cruz} et~al.}{2006}]{ambrociocruzetal2006}
{Ambrocio-Cruz} P.,  {Rosado} M.,    {de La Fuente} E.,  2006, \rmxaa, 42, 241

\bibitem[\protect\citeauthoryear{{Anders} \& {Grevesse}}{{Anders} \&
  {Grevesse}}{1989}]{abundances}
{Anders} E.,  {Grevesse} N.,  1989, \gca, 53, 197

\bibitem[\protect\citeauthoryear{{Arnaud}}{{Arnaud}}{1996}]{xspec}
Arnaud K. A., 1996, in Jacoby G. H., Barnes J., eds, 
Astronomical Data Analysis Software and Systems V Vol. 101 of Astronomical Society of the Pacific Conference Series, XSPEC: The First Ten Years. p. 17



\bibitem[\protect\citeauthoryear{{Bernitt}, {Brown}, {Rudolph},
  {Steinbr{\"u}gge}, {Graf}, {Leutenegger}, {Epp} \& {Eberle}}{{Bernitt}
  et~al.}{2012}]{bernittetal2012}
{Bernitt} S.,  {Brown} G.~V.,  {Rudolph} J.~K.,  {Steinbr{\"u}gge} R.,  {Graf}
  A.,  {Leutenegger} M.,  {Epp} S.~W.,    {Eberle} 2012, \nat, 492, 225

\bibitem[\protect\citeauthoryear{{Broersen}, {Vink}, {Kaastra} \&
  {Raymond}}{{Broersen} et~al.}{2011}]{broersenetal2011}
{Broersen} S.,  {Vink} J.,  {Kaastra} J.,    {Raymond} J.,  2011, \aap, 535,
  A11

\bibitem[\protect\citeauthoryear{{Cash}}{{Cash}}{1979}]{cash}
{Cash} W.,  1979, \apj, 228, 939

\bibitem[\protect\citeauthoryear{{Cox}, {Shelton}, {Maciejewski}, {Smith},
  {Plewa}, {Pawl} \& {R{\'o}{\.z}yczka}}{{Cox} et~al.}{1999}]{coxetal1999}
{Cox} D.~P.,  {Shelton} R.~L.,  {Maciejewski} W.,  {Smith} R.~K.,  {Plewa} T.,
  {Pawl} A.,    {R{\'o}{\.z}yczka} M.,  1999, \apj, 524, 179

\bibitem[\protect\citeauthoryear{{Dubner}, {Giacani}, {Goss} \&
  {Winkler}}{{Dubner} et~al.}{1994}]{dubneretal1994}
{Dubner} G.~M.,  {Giacani} E.~B.,  {Goss} W.~M.,    {Winkler} P.~F.,  1994,
  \aj, 108, 207

\bibitem[\protect\citeauthoryear{{Giacani}, {Dubner}, {Cappa} \&
  {Testori}}{{Giacani} et~al.}{1998}]{giacanietal1998}
{Giacani} E.~B.,  {Dubner} G.,  {Cappa} C.,    {Testori} J.,  1998, \aaps, 133,
  61

\bibitem[\protect\citeauthoryear{{Giuliani}, {Tavani}, {Bulgarelli}, {Striani}
  \& {Sabatini}}{{Giuliani} et~al.}{2010}]{giulianietal2010}
{Giuliani} A.,  {Tavani} M.,  {Bulgarelli} A.,  {Striani} E.,    {Sabatini}
  2010, \aap, 516, L11

\bibitem[\protect\citeauthoryear{{Harrus}, {Hughes}, {Singh}, {Koyama} \&
  {Asaoka}}{{Harrus} et~al.}{1997}]{harrusetal1997}
{Harrus} I.~M.,  {Hughes} J.~P.,  {Singh} K.~P.,  {Koyama} K.,    {Asaoka} I.,
  1997, \apj, 488, 781

\bibitem[\protect\citeauthoryear{{Kaastra}, {Mewe} \&
  {Nieuwenhuijzen}}{{Kaastra} et~al.}{1996}]{SPEX}
{Kaastra} J.~S.,  {Mewe} R.,    {Nieuwenhuijzen} H.,  1996, in {K.~Yamashita \&
  T.~Watanabe} ed., UV and X-ray Spectroscopy of Astrophysical and Laboratory
  Plasmas {SPEX: a new code for spectral analysis of X-ray and UV spectra.}.
p.~411

\bibitem[\protect\citeauthoryear{{Kawasaki}, {Ozaki}, {Nagase}, {Inoue} \&
  {Petre}}{{Kawasaki} et~al.}{2005}]{kawasakietal2005}
{Kawasaki} M.,  {Ozaki} M.,  {Nagase} F.,  {Inoue} H.,    {Petre} R.,  2005,
  \apj, 631, 935

\bibitem[\protect\citeauthoryear{{Kawasaki}, {Ozaki}, {Nagase}, {Masai},
  {Ishida} \& {Petre}}{{Kawasaki} et~al.}{2002}]{kawasakietal2002}
{Kawasaki} M.~T.,  {Ozaki} M.,  {Nagase} F.,  {Masai} K.,  {Ishida} M.,
  {Petre} R.,  2002, \apj, 572, 897

\bibitem[\protect\citeauthoryear{{Lazendic} \& {Slane}}{{Lazendic} \&
  {Slane}}{2006}]{lazendicslane2006}
{Lazendic} J.~S.,  {Slane} P.~O.,  2006, \apj, 647, 350

\bibitem[\protect\citeauthoryear{{Long}, {Blair}, {Matsui} \& {White}}{{Long}
  et~al.}{1991}]{longetal1991}
{Long} K.~S.,  {Blair} W.~P.,  {Matsui} Y.,    {White} R.~L.,  1991, \apj, 373,
  567

\bibitem[\protect\citeauthoryear{{Lopez}, {Pearson}, {Ramirez-Ruiz}, {Castro},
  {Yamaguchi}, {Slane} \& {Smith}}{{Lopez} et~al.}{2013}]{lopezetal2013}
{Lopez} L.~A.,  {Pearson} S.,  {Ramirez-Ruiz} E.,  {Castro} D.,  {Yamaguchi}
  H.,  {Slane} P.~O.,    {Smith} R.~K.,  2013, \apj, 777, 145

\bibitem[\protect\citeauthoryear{{Masai}}{{Masai}}{1997}]{masai1997}
{Masai} K.,  1997, \aap, 324, 410

\bibitem[\protect\citeauthoryear{{Miceli}, {Bocchino}, {Decourchelle}, {Ballet}
  \& {Reale}}{{Miceli} et~al.}{2010}]{micelietal2010}
{Miceli} M.,  {Bocchino} F.,  {Decourchelle} A.,  {Ballet} J.,    {Reale} F.,
  2010, \aap, 514, L2

\bibitem[\protect\citeauthoryear{{Milne}}{{Milne}}{1979}]{milne1979}
{Milne} D.~K.,  1979, Australian Journal of Physics, 32, 83

\bibitem[\protect\citeauthoryear{{Ohnishi et al.}}{{Ohnishi et al.}}{2011}]{ohnishietal2011}
Ohnishi T., Koyama K., Tsuru T. G., Masai K., Yamaguchi H.,
Ozawa M., 2011, PASJ, 63, 527

\bibitem[\protect\citeauthoryear{{Rho} \& {Petre}}{{Rho} \&
  {Petre}}{1998}]{rhopetre1998}
{Rho} J.,  {Petre} R.,  1998, \apjl, 503, L167

\bibitem[\protect\citeauthoryear{{Rosado}}{{Rosado}}{1983}]{rosado1983}
{Rosado} M.,  1983, \rmxaa, 8, 59

\bibitem[\protect\citeauthoryear{{Saken}, {Long}, {Blair} \& {Winkler}}{{Saken}
  et~al.}{1995}]{sakenetal1995}
{Saken} J.~M.,  {Long} K.~S.,  {Blair} W.~P.,    {Winkler} P.~F.,  1995, \apj,
  443, 231

\bibitem[\protect\citeauthoryear{{Schneiter}, {de La Fuente} \&
  {Vel{\'a}zquez}}{{Schneiter} et~al.}{2006}]{schneiteretal2006}
{Schneiter} E.~M.,  {de La Fuente} E.,    {Vel{\'a}zquez} P.~F.,  2006, \mnras,
  371, 369

\bibitem[\protect\citeauthoryear{{Shelton}, {Cox}, {Maciejewski}, {Smith},
  {Plewa}, {Pawl} \& {R{\'o}{\.z}yczka}}{{Shelton}
  et~al.}{1999}]{sheltonetal1999}
{Shelton} R.~L.,  {Cox} D.~P.,  {Maciejewski} W.,  {Smith} R.~K.,  {Plewa} T.,
  {Pawl} A.,    {R{\'o}{\.z}yczka} M.,  1999, \apj, 524, 192

\bibitem[\protect\citeauthoryear{{Smith} \& {Hughes}}{{Smith} \&
  {Hughes}}{2010}]{smithhughes2010}
{Smith} R.~K.,  {Hughes} J.~P.,  2010, \apj, 718, 583

\bibitem[\protect\citeauthoryear{{Spitzer} Jr.}{{Spitzer}}{1981}]{spitzer1981}
{Spitzer} Jr. L.,  1981, {Physical processes in the interstellar medium.}

\bibitem[\protect\citeauthoryear{{Tao}}{{Tao}}{1995}]{tao1995}
{Tao} L.,  1995, \mnras, 275, 965

\bibitem[\protect\citeauthoryear{{Uchida}, {Koyama}, {Yamaguchi}, {Sawada},
  {Ohnishi}, {Tsuru}, {Tanaka}, {Yoshiike} \& {Fukui}}{{Uchida}
  et~al.}{2012}]{uchidaetal2012}
{Uchida} H.,  {Koyama} K.,  {Yamaguchi} H.,  {Sawada} M.,  {Ohnishi} T.,
  {Tsuru} T.~G.,  {Tanaka} T.,  {Yoshiike} S.,    {Fukui} Y.,  2012, \pasj, 64,
  141

\bibitem[\protect\citeauthoryear{{Uchiyama}, {Funk}, {Katagiri}, {Katsuta},
  {Lemoine-Goumard}, {Tajima}, {Tanaka} \& {Torres}}{{Uchiyama}
  et~al.}{2012}]{uchiyamaetal2012}
{Uchiyama} Y.,  {Funk} S.,  {Katagiri} H.,  {Katsuta} J.,  {Lemoine-Goumard}
  M.,  {Tajima} H.,  {Tanaka} T.,    {Torres} D.~F.,  2012, \apjl, 749, L35

\bibitem[\protect\citeauthoryear{{Vink}}{{Vink}}{2012a}]{vink2012}
{Vink} J.,  2012a, \aapr, 20, 49

\bibitem[\protect\citeauthoryear{{Vink}}{{Vink}}{2012b}]{vinkreview}
{Vink} J.,  2012b, \aapr, 20, 49

\bibitem[\protect\citeauthoryear{{White} \& {Long}}{{White} \&
  {Long}}{1991}]{whitelong1991}
{White} R.~L.,  {Long} K.~S.,  1991, \apj, 373, 543

\bibitem[\protect\citeauthoryear{{Winkler}, {Olinger} \&
  {Westerbeke}}{{Winkler} et~al.}{1993}]{winkleretal1993}
{Winkler} P.~F.,  {Olinger} T.~M.,    {Westerbeke} S.~A.,  1993, \apj, 405, 608

\bibitem[\protect\citeauthoryear{{Yoshita}, {Tsunemi}, {Miyata} \&
  {Mori}}{{Yoshita} et~al.}{2001}]{yoshitaetal2001}
{Yoshita} K.,  {Tsunemi} H.,  {Miyata} E.,    {Mori} K.,  2001, \pasj, 53, 93

\end{thebibliography}
\end{document}